# SOLVING THE DE PRONY'S PROBLEM OF SEPARATION OF THE OVERLAPPING EXPONENTS IN DLTS

HOANG NAM NHAT

Faculty of Physics, University of Natural Sciences, VNU-HN.
334 Nguyen Trai, Thanh Xuan, Hanoi, Vietnam

ABSTRACT: This paper presents the solution to the De Prony's problem of separation of the overlapping exponents using the binomial coeficient as the weighting factors. The algebraic structure of the signal classes is discussed and the applicability of method is demonstrated.

## I. INTRODUCTION

The problem of separation of the overlapping exponents is old in physics and is important in both mathematics and physics, both theory and practice. In general we consider a measured signal of the form:

$$S = \sum_{i=1}^{m} C_i e^{-\omega_i t} + S_0 \qquad (1)$$

where $\omega_i$ is the unknown exponent (oscilating frequency, emission factor etc.), $t$ is the measured time, $C_i$ is the scale constant and $S_0$ is the back ground. In some cases, $\omega_i$ occurs as the constants (e.g. frequencies) but in general they depend on the external variables, e.g. on the temperature $T$. Where $\omega_i = f(T)$, S splits into a set of the decay-curves $S_i(T)$, each of them is recorded at certain fixed T. The problem is then to find the number $m$ of the exponents $\omega_i$, to estimate them and where applicable, to reveal the associated physical quantities (e.g. activation energy, concentration etc.). In 1795 de Prony proposed an algebraic method to solve the case when all $\omega_i$ are constant [1]. This method is still in use today and many alterations exist [2-4]. However, due to the lack of a priori knowledge of $m$ and to the low accuracy in occurence of noise, it can not succeed in many important situations. For the same reasons, the other newer methods involving least square fitting, such as the Fourier and the Laplace techniques [5-6], also fail frequently. The semiconductor physics met this problem again in the midle 1970s when examining the capacitance decays at the p-n junctions and found that when $\omega_i = f(T)$ none of the known methods can be used. A new method called the Deep Level Transient Spectroscopy (DLTS) has been introduced by D. Lang in 1974 for treating the deep level transients data [7]. Suppose the capacitance decay of the form (1), by placing two boxcar gates at time $t_1$ and $t_2$ Lang was able to construct a signal ($C_i=C_0$):

$$S(t_1) - S(t_2) = C_0 \sum_{i=1}^{m} \left( e^{-\omega_i t_1} - e^{-\omega_i t_2} \right) \qquad (2)$$

which shows maxima as $\omega_i$ varies according to the stepping temperature. The exact $\omega_i(T)$ were then determined from the positions of the maxima so avoiding the least square fitting. The DLTS provides spectroscopic visualisation to the number $m$ of actual exponential components. It underwent many revisions [8-12] (e.g the Fourier DLTS [8] has been commercially implemented in the Bio-Rad D5000 DLTS system) and was widely accepted as one of the most precise characterisation methods for semiconductors. The effort to extend this approach into the other areas suffers unfortunately two strong limitations: first, it requires ultimately $\omega_i = f(T)$ so is not applicable where $\omega_i$ holds contant; and second, it has poor separation resolution due to the Gaussian-shape spreading peak width. Any method, worth further consideration, should remove these two obstructions. This paper presents a method using a signal based on binomial series, $P_n^{(k)}/C_0 = X^k(1-X)^n$; we show that (*i*) its peak width depends inversely on the number of digits that input data may be accurately determined (so offering the separation resolution that is theoretically unbound); (*ii*) it is spectroscopic for both cases $\omega_i = f(T)$ and $\omega_i = const$, so providing the visual determination to the number *m* of the exponential components. The algebraic structure of this signal class is also discussed in details.

## II. THE BINOMIAL SIGNAL FORM

Suppose one component case and that the input data were collected in the equal time interval $\Delta t$. The *j*-th successive value of S, starting from some given $t_0=k\Delta t$, takes form:

$$S(t_0 + j\Delta t) = C_0 e^{-\omega t_0}\left(e^{-\omega \Delta t}\right)^j + S_0 \qquad (3)$$

For $j = 0...n$, multiply each S by a constant $a_j^{(n)}$ defined as a binomial coefficient:

$$a_j^{(n)} = (-1)^j C_j^n = (-1)^j \frac{n!}{j!(n-j)!} \qquad (4)$$

We call a *binomial signal order n and position k* a function $P_n^{(k)}$ defined as:

$$P_n^{(k)} = C_0 \sum_{j=0}^{n} a_j^{(n)} e^{-\omega t_0}\left(e^{-\omega \Delta t}\right)^j + S_0 \sum_{j=0}^{n} a_j^{(n)} \qquad (5)$$

Since for the binomial coeficients $\sum_{j=0}^{n} a_j^{(n)} = 0$, the $P_n^{(k)}$ strips down the back ground $S_0$ immediately.

Substitute $t_0=k\Delta t$ and denote $X = e^{-\omega \Delta t}$, it follows according to the binomial theorem:

$$P_n^{(k)} = C_0 \sum_{j=0}^{n} a_j^{(n)} X^{k+j} = C_0 X^k (1-X)^n \qquad (6)$$

For $n=1$, this function implies the Lang's signal (2). It is definitely positive continuous function for $X \in \langle 0,1 \rangle$ (i.e. for $t \in (0,\infty)$) and its first order derivation according to $X$ (and to $\omega$ too) reveals the position of the maximum:

$$X_{max} = \frac{k}{n+k} \qquad (7)$$

For $k=n$ ($X=1/2$) i.e. the graph of $P_n^{(n)}$ is symmetric, for $k<n$ ($X<1/2$) the graph moves to the left-side; and for $k>n$ ($X>1/2$) inversely. By substituting (7) into (6) we found the peak height:

$$Y_n^{(k)} = Max\, P_n^{(k)} = C_0 \left(\frac{k}{n+k}\right)^k \left(\frac{n}{n+k}\right)^n \qquad (8)$$

This value decreases very fast for large *n* and *k*, as can easily be seen for $k=n$: $Y_n^{(n)} = C_0/2^{2n}$. The peak width at half maximum $\delta_n^{(k)}$ can be found by solving the equation $P_n^{(k)} = Y_n^{(k)}/2$. To prove that $\delta_n^{(k)} \to 0$ when $(n+k)\to\infty$, we derive $\delta_n^{(k)}$ for $n=k$. For this the equation $P_n^{(k)} = Y_n^{(k)}/2$ reduces to $X^2 - X + \sqrt[n]{C_0}/2^{2+1/n} = 0$ which has the separation between its two roots:

$$\delta_n^{(n)} = X_2 - X_1 = \sqrt{1 - \frac{4\sqrt[n]{C_0}}{2^{2+1/n}}} \qquad (9)$$

(whereas $X_2+X_1=1$). It is evidently that $\lim_{n\to\infty} \delta_n^{(n)} = 0$. Next data show how fast $\delta_n^{(n)}$ evolves.

TABLE I. Development of $\delta_n^{(n)}$ according to $n$, $C_0=1$.

| $n$ | 1 | 2 | 3 | 4 | 5 | 6 | 7 | 8 | 9 |
|---|---|---|---|---|---|---|---|---|---|
| $\delta_n^{(n)}$ | 0.71 | 0.54 | 0.45 | 0.40 | 0.36 | 0.33 | 0.30 | 0.29 | 0.27 |

Creating the *m*-order binomial signals on basis of the existing ones yields *m* shift in the order or position:

$$\sum_{l=0}^{m} a_l^{(m)} P_n^{(k+l)} = P_{n+m}^{(k)} \tag{10}$$

$$\sum_{l=0}^{m} a_l^{(m)} P_{n+l}^{(k)} = P_n^{(k+m)} \tag{11}$$

whereas only the 2-order binomial signal composed on basis of the $\log\left(P_n^{(k)}\right)$ is non-zeroed:

$$L_m^{(k)} = \sum_{l=0}^{m} a_l^{(m)} \log\left(P_n^{(k+l)}\right) = \begin{pmatrix} -\log X & \text{for } m=2 \\ 0 & \text{for } m>2 \end{pmatrix} \tag{12}$$

This because the sum reduces to $\log X \sum_{l=0}^{m} a_l^{(m)} l$ whose sum is equal $-1$ for *m=2* and 0 for all other *m*.

Since $Y_n^{(k)}$ in (8) decreases for large (*n+k*) we construct a new signal having $1/Y_n^{(k)}$ as a scale constant:

$$Q_n^{(k)} = \frac{P_n^{(k)}}{Y_n^{(k)}} = \left(\frac{n+k}{k}\right)^k \left(\frac{n+k}{n}\right)^n X^k (1-X)^n \tag{13}$$

This signal strips off $C_0$, has 1 as maximum and shows maximum also according to *k* as variable, the position of the maximum is:

$$k_{max} = \frac{nX}{1-X} \tag{14}$$

which corresponds to the same point as specified in (7). The density function for binomial distribution may also be derived from $P_n^{(k)}$, denote *m=n+k*:

$$p(m,k) = \frac{C_k^m}{C_0} P_n^{(k)} = C_k^m X^k (1-X)^{m-k} \tag{15}$$

which has the mean $<k>=mX$ (same as (7)) and the variance $mX(1-X)$.

Now suppose *m* overlapping centers. $P_n^{(k)}$ will be:

$$P_n^{(k)} = \sum_{i=1}^{m} C_i X_i^k (1-X_i)^n \tag{16}$$

where $X_i$ stands for $e^{-\omega_i \Delta t}$. With assumption that all $\omega_i$ are independent, i.e. $\frac{d}{dX_i}\left\{\sum_{j=1}^{m} X_{j\neq i}^k (1-X_{j\neq i})^n\right\} = 0$, the $P_n^{(k)}$ shows maximum according to $X_i$ exactly when this $X_i$ scans through the value $k/(n+k)$. The same is for $Q_n^{(k)}$ which should show the corresponding maximum for $X_i$ when *k* passes through $nX_i/(1-X_i)$. This fact is extremely important since it implies that all maximum positions are independent and do not alter each other.

To explain the algebraic structure of the (S, *t*) plane with respect to $P_n^{(k)}$, suppose a fixed time position *t=W/b*, where *b* is an arbitrary constant and *W* is a period width. A $P_n^{(k)}$ signal that involves *n/2* measured data left from *t* and *n/2* data right from it, depends on the fineness of the steps $\Delta t$. Denote the first index $k_x$, from (7) it follows:

$$\omega_{max} = \frac{1}{\Delta t}\log\left(\frac{n+k_x}{k_x}\right) = \log\left(1+\frac{1}{k_x/n}\right)^{1/\Delta t} \quad (17)$$

If $N$ is the total number of collected data $N=W/\Delta t$, then $k_x = N/b - n/2$. Substitute this into (17) and use the Euler's formula $\lim_{n\to\infty}(1+1/n)^n = e$ it follows that:

$$\lim_{\Delta t\to 0}\omega_{max} = \lim_{\Delta t\to 0}\log\left(1+\frac{1}{t/n\Delta t - 1/2}\right)^{1/\Delta t} = \frac{n}{t} \quad (18)$$

For $n=1$ (Lang's signal) this implies that $\omega_{max}\to 1/t$ as was observed in [13]; for $n=m>1$, $\omega_{max}\to m/t$. Now consider the moving of $t$ to $t/a$, for $a=m/n$. According to (18) the limit for $\omega_{max}$ at this time is $an/t=m/t$ which is just the value given by $P_m^{(k)}$ at time $t$. So to each point $s(S, t)$ in the $(S, t)$ plane one value $\omega(n,t) = \lim_{\Delta t\to 0}\omega_{max}$ may be attached:

$$\omega(n,t) = \omega(1,t/n) = n\omega(1,t) \quad (19)$$

By insertion of (18) into (1), the signal S becomes:

$$S^n = e^{-n}\sum_{i=1}^{m}C_i + S_0 \quad (20)$$

Providing this $y=S^n$ as a horizontal line and $x=t$ as a vertical line, the intersection of the two determines a point lying on the curve $S_i(T)$ whose exponent is just $\omega(n,t)$. This completely describes $(S, t)$: with respect to $P_n^{(k)}$ all points $s(S,t)$ are equivalent! There are many signal forms whose orders are the real numbers, not only the integers as here for $P_n^{(k)}$ [13].

### III. APPLICATION OF THE METHOD

The use of the signals of higher orders $n$ depends strictly on the number $N$ of the decimal digits that the inputs may be determined. Suppose that the last $N$-$th$ digit corresponds to the error, i.e. the error is of $10^{-N}$ order, the signal-to-noise ratio (SNR) evolves as $Y_n^{(k)}/10^{-N} \approx C_0 4^{-n}/10^{-N} \cong C_0 10^{N-0.602n}$. The table below illustrates this dependence on $n$ for $N=3$.

TABLE II. Dependence of SNR on $n$ for N=3 ($C_0$=1).

| $n$ | 1 | 2 | 3 | 4 | 5 | 6 | 7 | 8 |
|---|---|---|---|---|---|---|---|---|
| SNR | 250.0 | 62.5 | 15.6 | 3.9 | 0.98 | 0.24 | 0.06 | 0.02 |

The good estimation for the maximal 'error free' signal order is $n_{max}=1.6N$.

For semiconductor physics, this method provides the substantial improvement in detection of the closely spaced deep states. Fig.1 shows the consideration for separation of the two states with activation energy 0.45eV and 0.43eV by a series of the $P_2^{(5)}, P_3^{(3)}$ and $P_5^{(1)}$ signals. Such small separations are practically impossible to identify by the recent techniques.

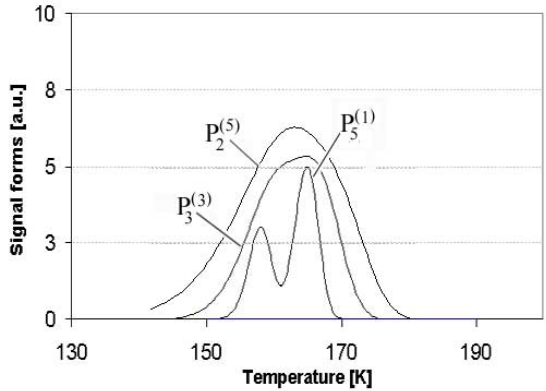 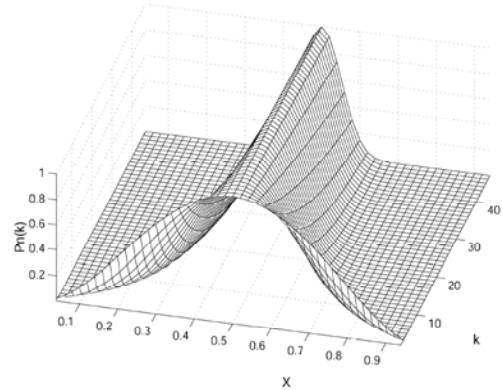

Theoretically one may even separate two physically separated clusters with equal activation energy but different concentration using the signals of higher order, such as $P_{15}^{(k)}$, if the physics of emission follows (1) and if the transients can be measured enough accurately, i.e. up to $N=n_{max}/1.6 = 15/1.6 > 9$ decimal digits.

REFERENCES
[1] de Prony, Baron Gaspard Riche (1795). *Journal de l'École Polytechnique*, Vol.**1**, cahier 22, 24-76.
[2] Michael E. Wall, Andreas Rechtsteiner and Luis M. Rocha1, "*Singular value decomposition and principal component analysis*" In "A Practical Approach to Microarray Data Analysis" edited by D.P. Berrar, W. Dubitzky and M. Granzow, Kluwer: Norwell, MA, 2003. pp. 91-109. LANL LA-UR-02-4001.
[3] S. Lawrence Marple, Jr., "*Digital Spectral Analysis with Applications*", Prentice-Hall, 1987, p.303-349
[4] Osborne, M.R. and Smyth, G.K. (1991). *SIAM Journal of Scientific and Statistical Computing*, **12**, 362-382.
[5] J. Morimoto, M. Fudamoto, K. Tahira, T. Kida, S. Kato, T. Miyakawa, *Jap. J. Appl. Phys*. **26** (10) (1987) p.1634-1640.
[6] M. Okuyama, H. Takakura, Y. Hamakawa, *Solid-State Elect.*, **26** (1983) p.689-694.
[7] D.V. Lang, *J. Appl. Phys*. **45** (1974) p.3023.
[8] S. Weiss & R. Kassing, *Solid State Electronics*, Vol. 31, **12** (1988) p. 1733
[9] L. Dobaczewski, P. Kaczor, I.D. Hawkins, A.R.Peaker, *Mat.Sci.and Tech.* **11** (1994) p. 194-198.
[10] I. Thurzo, D. Pogany, K. Gmucova, *Solid-State Elect.*, **35** (1992) p.1737-1743.
[11] Z. Su and J.W. Farmer, *J. Appl. Phys*. **68** (1990) p.4068-4070
[12] F. R. Shapiro, S.D. Senturia and D. Adler, *J. Appl. Phys.* **55**((1984) p. 3453-3459.
[13] Hoang Nam Nhat and Pham Quoc Trieu, VNU *Journal of Sciences, Mathematics-Physics.*, Vol. XVIII, No.4, 2002, p.28-36